# Gate-Induced Interfacial Superconductivity in 1T-SnSe$_2$

Junwen Zeng[1], Erfu Liu[1], Yajun Fu[1,2], Zhuoyu Chen[3,4], Chen Pan[1], Chenyu Wang[1], Miao Wang[1], Yaojia Wang[1], Kang Xu[1], Songhua Cai[5], Xingxu Yan[5], Yu Wang[1], Xiaowei Liu[1], Peng Wang[5], Shi-Jun Liang[1*], Yi Cui[3,4], Harold Y. Hwang[3,4], Hongtao Yuan[1,3,4*] & Feng Miao[1*]

[1] National Laboratory of Solid State Microstructures, School of Physics, Collaborative Innovation Center of Advanced Microstructures, Nanjing University, Nanjing 210093, China.

[2] School of Material Science and Engineering, Southwest University of Science and Technology, Mianyang, China

[3] Geballe Laboratory for Advanced Materials, Stanford University, Stanford, California 94305, USA.

[4] Stanford Institute for Materials and Energy Sciences, SLAC National Accelerator Laboratory, Menlo Park, California 94025, USA.

[5] College of Engineering and Applied Sciences, Nanjing University, Nanjing 210093, China.

*Corresponding authors
 miao@nju.edu.cn (F.M.);
 htyuan@nju.edu.cn (H.T.Y.);
 sjliang@nju.edu.cn (S.J.L.).

**ABSTRACT:** Layered metal chalcogenide materials provide a versatile platform to investigate emergent phenomena and two-dimensional (2D) superconductivity at/near the atomically thin limit. In particular, gate-induced interfacial superconductivity realized by the use of an electric-double-layer transistor (EDLT) has greatly extended the capability to electrically induce superconductivity in oxides, nitrides and transition metal chalcogenides and enable one to explore new physics, such as the Ising pairing mechanism. Exploiting gate-induced superconductivity in various materials can provide us with additional platforms to understand emergent interfacial superconductivity. Here, we report the discovery of gate-induced 2D superconductivity in layered 1T-SnSe$_2$, a typical member of the main-group metal dichalcogenide (MDC) family, using an EDLT gating geometry. A superconducting transition temperature $T_c \approx$ 3.9 K was demonstrated at the EDL interface. The 2D nature of the superconductivity therein was further confirmed based on 1) a 2D Tinkham description of the angle-dependent upper critical field $B_{c2}$, 2) the existence of a quantum creep state as well as a large ratio of the coherence length to the thickness of superconductivity. Interestingly, the in-plane $B_{c2}$ approaching zero temperature was found to be 2-3 times higher than the Pauli limit, which might be related to an electric field-modulated spin-orbit interaction. Such results provide a new perspective to expand the material matrix available for gate-induced 2D superconductivity and the fundamental understanding of interfacial superconductivity.

Exploration of 2D superconductivity in new material systems has attracted intensive research interest in condensed matter physics.[1-4] In particular, EDL gating has been utilized as a powerful technique to realize interfacial carrier accumulation at an ultra-high density level for realization of 2D superconductivity in various semiconductors and insulators, such as transition metal dichalcogenides (TMDCs)[5-13], ZrNCl[14,15], and perovskite oxides (e.g., $SrTiO_3$ and $KTaO_3$).[16,17] In these systems, emergent interfacial phenomena have been observed, such as the 2D Ising electron pairing protected by spin-valley locking in $MoS_2$, which induces the enhancement of the Pauli paramagnetic limit[8,9], and the suppression of charge density waves in favor of superconductivity in $TiSe_2$, providing an electrical method to control the many-body interaction.[7]

Compared with TMDCs that show rich electronic properties typically due to the versatile electron filling of heavy $d$ orbitals of the transition metals,[18-20] the electronic structures of main-group MDCs can be quite different.[21-24] For example, a typical main-group layered semiconductor, InSe, with conduction and valence band edges mainly contributed by $s$- and $p$- orbital electrons, was reported to host a small effective mass and high electron mobility, leading to demonstration of the quantum Hall state.[25,26] Previous studies of electric field-tuned layered main-group MDCs have focused on normal-state properties, such as the insulator-metal transition in $SnS_2$[27] and the quantum Hall effect in InSe.[25] Gating-induced 2D superconductivity in the main-group MDCs has not yet been observed. $SnSe_2$, one of main-group MDCs materials, has been reported to support superconducting state by organometallic intercalation.[28-32] The question about whether its intrinsic 2D superconductivity in the few-layer $SnSe_2$ exists is worth to be explored.

In this study, we demonstrated a gate-induced superconductivity with $T_c \approx 3.9$ K in mechanically exfoliated 1T-$SnSe_2$ based on the ionic liquid gating technique. Angle and temperature-dependent $B_{c2}$ measurements are used to show the 2D nature of the superconducting state. The temperature dependence of the resistance in out-of-plane magnetic fields reveals the nature and dynamics of the vortex state below $T_c$. The in-plane $B_{c2}$ measurements suggest a spin-orbit interaction (SOI)-related depairing mechanism.

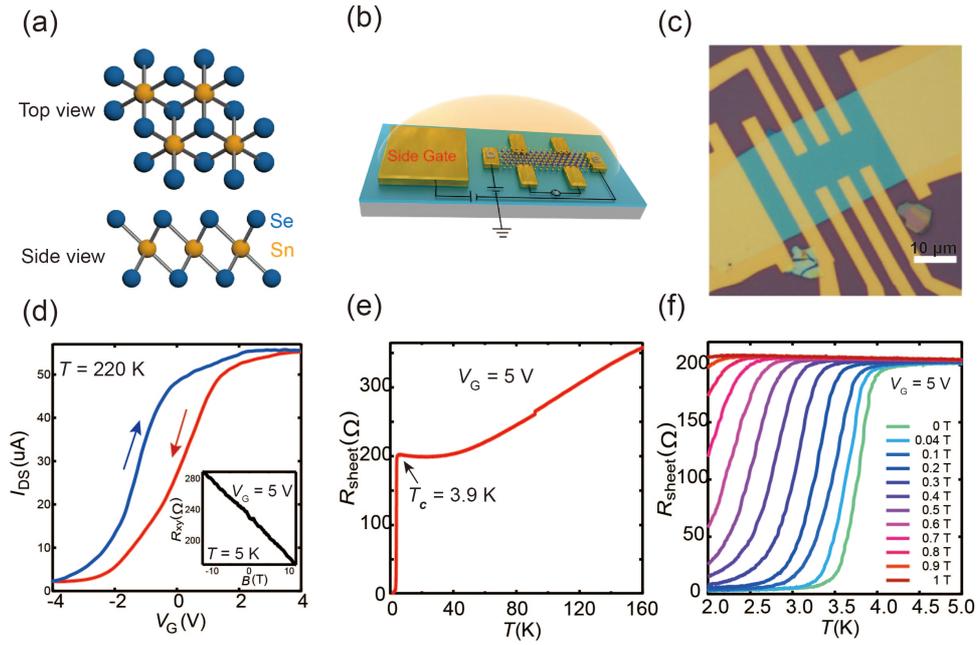

**Figure 1.** Crystal structure of 1T-SnSe$_2$, transfer curve and superconductivity of SnSe$_2$-EDLT. (a) The top and side views of the crystal structure. (b) Sketch of a SnSe$_2$-EDLT device. (c) Optical microscope image of a typical multi-terminal Hall bar device. The scale bar is 10 μm. (d-f) Transfer curve and superconductivity of dev#01. (d) $I_{sd}$ as a function of sweeping $V_G$, with $V_{sd}$ = 0.1 V. The inset shows the magnetic field dependence of $R_{xy}$ at $T$ = 5 K with $V_G$ = 5 V. (e) Temperature-dependent resistance between 2 and 160 K with $V_G$ = 5 V. A metal-to-superconductor transition occurs at $T_c$ = 3.9 K. (f) Temperature dependence of the magnetoresistance. The measurements were carried out using an excitation current of 50 nA and varying out-of-plane magnetic fields up to 1 T under the same $V_G$.

The SnSe$_2$ studied here is a layered semiconductor with a 1T-type crystal structure.[24] As shown in Fig. 1a (with top and side views), the Sn layer is sandwiched between two Se layers in monolayer 1T-SnSe$_2$, with ABC (Se-Sn-Se) atomic stacking. A Sn atom is surrounded by six Se atoms, forming an octahedral coordination.[33] Micro-Raman spectroscopy and transmission electron microscopy (TEM) were employed to understand the crystalline properties of the mechanically exfoliated thin flakes. The Raman spectra measured using a 514 nm laser at room temperature showed two sharp peaks, as shown in Fig. S1c (the out-of-plane $A_{1g}$ mode located at 184 cm$^{-1}$ and in-plane $E_g$ mode located at 110 cm$^{-1}$). Similar to other reports, such an observation indicates a

1T-type crystal phase for the SnSe$_2$ single crystals.[34,35] The high-angle annular dark-field scanning transmission electron microscopy (HAADF-STEM) image is shown in Fig. S1d; the observed arrangement of Sn and Se atoms agrees well with the top view shown in Fig. 1a.

Fig. 1b shows a schematic illustration of a SnSe$_2$-EDLT device with an ionic liquid gate (DEME-TFSI). The side gate drives corresponding ions towards the channel surface, resulting in the formation of an EDL capacitance with an ultrathin (~1 nm) Helmholtz layer.[15,17,19] As a result, a high carrier density can be accumulated at the surface of the thin flakes. Fig. 1c shows the optical image of a typical SnSe$_2$ device (see fabrication details in methods section). We first measured the transfer characteristics of the SnSe$_2$-EDLT at 220 K (above the glass transition temperature of the ionic liquid) with a current ON/OFF ratio of ~ 25, as shown in Fig. 1d.[5,9,36] The application of a side gate voltage of $V_G$ = 5 V leads to the build-up of a sheet carrier density as high as 1.4 × 10$^{14}$ cm$^{-2}$ at $T$ = 5 K (inset of Fig. 1d). In this paper, the data of superconductivity are from four typical devices. Note that the ionic liquid gating technique may induce inhomogeneous carrier doping across the surface of SnSe$_2$ samples, leading to subtle different superconductivity performance in different geometric devices, as pointed out in the prior work.[6]

The temperature-dependent sheet resistance curve ($R_{sheet}$-$T$) of a typical device with $V_G$ = 5 V is shown in Fig. 1e. The data show a metallic state during cooling to 4 K, with a sharp drop in resistance observed at temperatures below 4 K that is associated with the emergence of superconductivity. We estimated the critical temperature $T_c$ to be approximately 3.9 K ($T_c$ is defined as the temperature at which $R_{sheet}$ is equal to 90% × $R_n$, where $R_n$ is the normal-state resistance). Fig. 1f shows the $R_{sheet}$-$T$ curves for varying out-of-plane magnetic fields. $T_c$ deceases with increasing magnetic field until the superconducting state is completely suppressed at approximately 1 T. Out-of-plane magnetoresistance measurements at various temperatures and the $V$-$I$ curves measured under different out-of-plane magnetic fields are shown in the supplementary information. At higher temperatures, the superconductivity can be suppressed by a smaller magnetic field. The color plot of $V$ versus $I$ and $B$ shows that the critical current is diminished by the increasing magnetic field (details shown in Fig. S3).

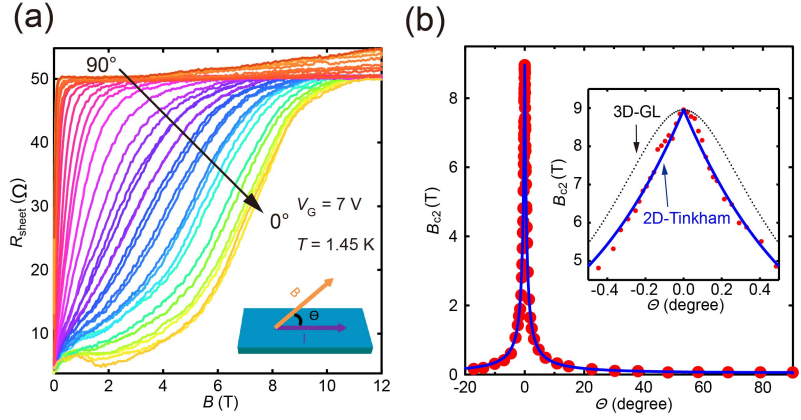

**Figure 2.** Confirmation of 2D superconductivity. (a) The magnetoresistance of dev #02 measured at different angles at $T$ = 1.45 K and $V_G$ = 7 V. The inset shows a schematic description of the angle $\theta$ between magnetic field and flake surface. (b) The $B_{c2}$ as a function of $\theta$ extracted from (a). The inset shows $B_{c2}$ (symbols) plotted in the range $\theta = \pm 0.5°$. The blue solid line and the black dashed line are fits to the 2D Tinkham formula and 3D Ginzburg-Landau anisotropic model, respectively. The $B_{c2}$ curve agrees well with the 2D Tinkham model.

Angle-dependent magnetoresistance measurements were also carried out to confirm the 2D nature of the observed superconducting state. Fig. 2a shows the magnetoresistance (with $T$ = 1.45 K and applied gate voltage $V_G$ = 7 V) at various $\theta$ (defined as the angle between the magnetic field and flake surface, as illustrated in the inset of Fig. 2a); the magnetoresistance is highly $\theta$-dependent. Here, the magnetoresistance at approximately $\theta = 0°$ was found to decrease with increasing magnetic field between 1 and 2 T. This observation is similar to a previous study of the interface of LaAlO$_3$/SrTiO$_3$ and amorphous Pb films that showed enhancement of superconductivity induced by an in-plane magnetic field over a small range, which may be related to the strong SOI.[37] We defined $B_{c2}$ to be the critical magnetic field at which the resistance reaches 90% × $R_n$, and plotted the measured $B_{c2}$ versus $\theta$ in Fig. 2b. $B_{c2}$ is approximately 0.06 T for $\theta$ = 90° and reaches a value as high as 9.0 T for $\theta$ = 0°. We fitted the $B_{c2}$-$\theta$ curves using both the 2D Tinkham formula, $\left(\frac{B_{c2}(\theta)\cos(\theta)}{B_{c2}^{\parallel}}\right)^2 + \left|\frac{B_{c2}(\theta)\sin(\theta)}{B_{c2}^{\perp}}\right| = 1$, and three-

dimensional (3D) Ginzburg-Landau anisotropic model, $\left(\frac{B_{c2}(\theta)\cos(\theta)}{B_{c2}^{\parallel}}\right)^2 +$ $\left(\frac{B_{c2}(\theta)\sin(\theta)}{B_{c2}^{\perp}}\right)^2 = 1$ (where $B_{c2}^{\parallel}$ and $B_{c2}^{\perp}$ are the upper critical fields at $\theta = 0°$ and 90°, respectively).[8,9,14] The inset of Fig. 2b clearly shows that the fitting curve agrees well with the 2D Tinkham model, indicating the 2D nature of the observed superconductivity.

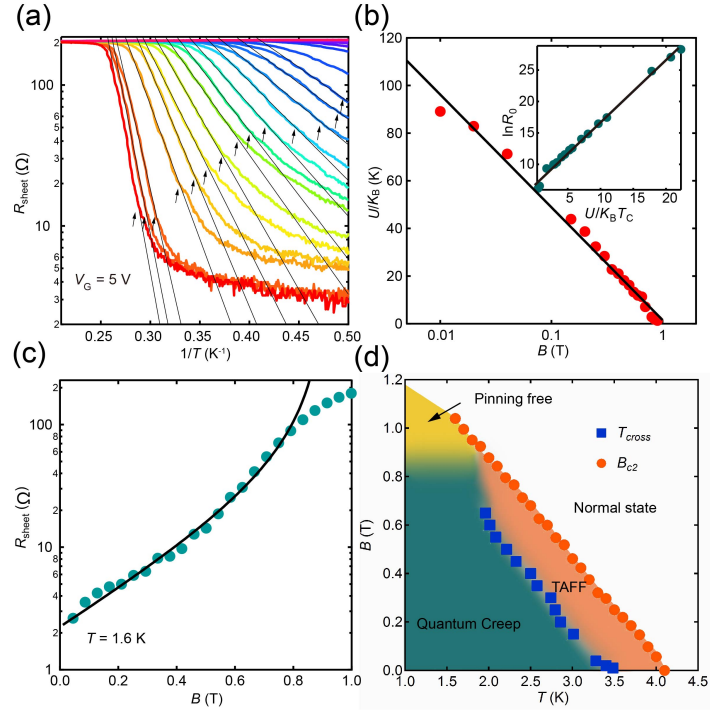

**Figure 3.** Vortex dynamics analysis. (a) An Arrhenius plot for the temperature dependence of the sheet resistance measured at different magnetic fields based on dev#01 at $V_G = 5$ V, which indicates thermally activated behavior (black solid lines). (b) Magnetic field dependence of the activation energy ($U(B)/k_B$), plotted on a semi-logarithmic scale. The activation energy is well described by the equation $U(B) = U_0\ln(B_0/B)$ (the black solid line). The inset shows the curve of $\ln R_0$ versus $U(B)/k_B T_c$. (c) The measured sheet resistance as a function of magnetic field at 1.6 K, which is fitted well by a quantum creep model below $B = 0.8$ T (the solid black line). (d) Vortex phase diagram of the SnSe$_2$-EDLT device. The solid circles show $B_{c2}$ at different temperatures. The solid squares $T_{cross}$ are derived from the crossing points of the $R_{sheet}$ - $1/T$ curves with the thermal activation fitting curves shown in (a) (black arrows).

To understand the microscopic vortex state in the 2D superconductor,[14,38,39] we plot the $R_{\text{sheet}}$-$T$ curves at different out-of-plane magnetic fields $B$ (Fig. 1f) using the Arrhenius convention in the Fig. 3a. We note that for temperatures immediately below the onset of the resistance drop, the observed $R_{\text{sheet}}$-$T$ behavior is consistent with a thermally activated process: the temperature is comparable to the finite energy barriers expected for the motion of vortices, with the vortices participating in a thermally activated flux flow (TAFF).[40,41] The black lines are fits to a model of thermally activated behavior described by $R_{\text{sheet}} = R_0(B)\exp(-U(B)/k_B T)$, where $U(B)$ is the activation energy and $k_B$ is the Boltzmann constant. The magnetic-field dependence of the extracted $U(B)/k_B$ is plotted in Fig. 3b, while $R_0(B)$ as a function of $U(B)/k_B T_c$ is plotted in the inset. These two curves can be well fitted using the equations $U(B) = U_0 \ln(B_0/B)$ and $\ln(R_0(B)) = U(B)/k_B T_c + A$ (A is a constant), respectively, where $U_0$ is the vortex-antivortex binding energy. From the $U(B)$ fitting, we obtained $U_0/k_B = 20.6$ K and $B_0 = 1.1$ T, which are comparable to values obtained in the previous studies of ZrNCl and NbSe$_2$.[14,39]

Interestingly, the applied magnetic field gives rise to an apparent metallic ground state, as indicated by the saturating finite resistance as one approaches zero temperature. Note that this is different from the conventional magnetic field-induced superconductor-insulator transition in disordered systems.[42,43] We show that the magnetic field dependence of the metallic ground state resistance follows a quantum creep model, which describes the temperature-independent quantum tunneling of vortices in the following equations:[14,44]

$$R_{\text{sheet}} \sim \frac{\hbar}{4e^2} \frac{\mathcal{L}}{1-\mathcal{L}}, \qquad (1)$$

$$\mathcal{L} \sim \exp\left\{C \frac{\hbar}{e^2} \frac{1}{R_N} \left(\frac{B-B_{c2}}{B_{c2}}\right)\right\}, \qquad (2)$$

where $\hbar$ is the reduced Planck constant, $e$ is the elementary charge, and $C$ is a dimensionless constant. As shown in Fig. 3c, at $T = 1.6$ K, the magnetic field dependence of $R_{\text{sheet}}$ can be described by a quantum creep model below 0.8 T, indicating that the vortices move via quantum creep. We further plot the magnetic field-temperature vortex phase diagram in Fig. 3d. TAFF and quantum creep states are separated by $T_{\text{cross}}$ (annotated by blue squares), which is defined as the point (indicated by the black arrows in Fig. 3a) at which the thermal activation fitting curve deviates from the measured data. The system passes through a magnetic field driven superconductor-metal transition while approaching zero temperature.

To shed light on the unique characteristics of the electric field-induced 2D superconductivity in main-group MDCs, we also measured the temperature dependence of the magnetoresistance in both in-plane and out-of-plane magnetic fields. We can extract the coherence length and thickness of the superconductor by fitting the in-plane and out-of-plane $B_{c2}$ with the 2D Ginzburg–Landau model:

$$B_{c2}^{\perp}(T) = \frac{\Phi_0}{2\pi\xi_{GL}(0)^2}\left(1 - \frac{T}{T_c}\right), \qquad (3)$$

$$B_{c2}^{\parallel}(T) = \frac{\Phi_0\sqrt{12}}{2\pi\xi_{GL}(0)d_{sc}}\left(1 - \frac{T}{T_c}\right)^{\frac{1}{2}}, \qquad (4)$$

where $\Phi_0$ denotes the magnetic flux quantum ($2.07 \times 10^{-15}$ Wb), $\xi_{GL}(0)$ is the in-plane 2D GL coherence length at $T = 0$ K and $d_{sc}$ is the thickness of the superconductor. Based on the analysis shown in Fig. S5, we obtained $\xi_{GL}(0) \approx 53.7$ nm and $d_{sc} \approx 1.6$ nm. Note that the value for $d_{sc}$ is close to the thickness of bilayer SnSe$_2$ (the interlayer distance of SnSe$_2$ is 0.62 nm[34]) and $d_{sc} \ll \xi_{GL}(0)$, which clearly indicate the 2D nature of our system.

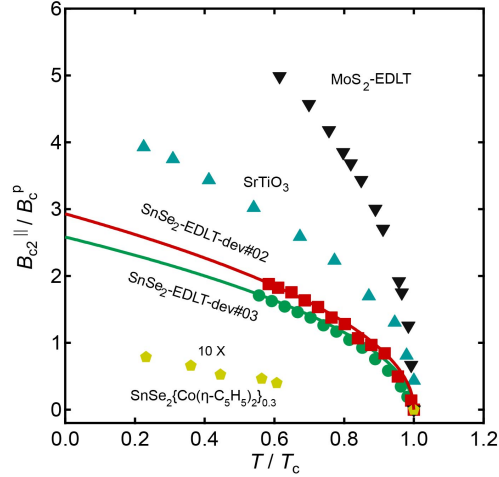

**Figure 4.** The temperature dependence of $B_{c2}^{\parallel}/B_c^p$ of two SnSe$_2$-EDLT devices (red square symbols (dev#02) and green circle symbols (dev#03)), MoS$_2$-EDLT (black triangles) (ref. 8), SrTiO$_3$ (blue triangles) (ref. 48), and SnSe$_2$\{Co($\eta$-C$_5$H$_5$)$_2$\}$_{0.3}$ (Yellow pentagons) (ref. 32). The values shown for SnSe$_2$\{Co($\eta$-C$_5$H$_5$)$_2$\}$_{0.3}$ were expanded tenfold. The red and green lines denote the fitting curves obtained from the 2D Ginzburg–Landau models.

We now discuss the pairing mechanism in this system. In a conventional superconductor, a cooper pair is composed of two electrons with opposite spin. In an external magnetic field, the superconducting state will be destroyed if the magnetic energy required for flipping the electron spin is comparable to the superconducting energy gap. The corresponding magnetic field is known as the Pauli paramagnetic limit $B_c^p$, which is given by $B_c^p \approx 1.85 \times T_c$ assuming the weak coupling limit of superconductivity.[45,46] If we plot the normalized $B_{c2}^\parallel/B_c^p$ as a function of temperature (Fig. 4) for the superconducting SnSe$_2$ devices, we can see that $B_{c2}^\parallel$ exceeds $B_c^p$ by a factor of 2-3 for temperatures approaching 0 K. The enhancement of the Pauli paramagnetic limit under an in-plane magnetic field has been found in many 2D superconductors with a strong SOI effect.[8,9,47,48] By application of a gate voltage and further breaking of the inversion symmetry, the SOI in SnSe$_2$ can be electrically modulated, as confirmed by magnetoresistance measurements (details shown in Supplementary Information Fig. S6). This also implies that the strong SOI induced by the perpendicular electric field plays an important role in the enhancement of the Pauli paramagnetic limit in SnSe$_2$. We also note that in bulk superconducting SnSe$_2$ with a lamellar organometallic intercalation, where the orbital effect dominates the pair-breaking mechanism of the superconductivity under a magnetic field, the $B_{c2}^\parallel$ is far smaller than the Pauli paramagnetic limit.[32] However, in the 2D limit, where $d_{sc} \ll \xi_{GL}(0)$, the orbital depairing effect is dramatically suppressed in the SnSe$_2$-EDLT. For comparison, we plotted the data reported for a MoS$_2$-EDLT[8] and SrTiO$_3$[48] as well in Fig. 4.

In summary, we have induced a superconducting state below $T_c$ = 3.9 K in SnSe$_2$ by using the ionic liquid gating technique. The 2D nature of the superconducting state is confirmed by both the angle-dependent upper critical field $B_{c2}$ described by 2D Tinkham and the fact that $d_{sc} \ll \xi_{GL}(0)$. The in-plane $B_{c2}$ of the devices greatly exceeds the Pauli limit, which might be associated with an SOI modulated by the strong perpendicular electric field. Note that the occurrence of superconductivity in 1T-SnSe$_2$ induced by an EDL expands the scope for 2D superconductors, and further paves the way for investigating new and undiscovered superconductors.

# Methods

## Materials and devices

We used the standard mechanical exfoliation method to obtain $SnSe_2$ flakes on 285-nm-thick $SiO_2$ wafers. The thickness of the flakes was confirmed by a Bruker Multimode 8 atomic force microscope (AFM). By using electron-beam lithography (a pattern generation system NPGS with a scanning electron microscope FEI F50), the pre-selected flake was patterned into a multi-terminal Hall bar configuration (as shown in Fig. 1b), followed by the deposition of Ti/Au (5/50 nm) metal contacts by standard electron beam evaporation.

## Transport measurements

Following the nano-fabrication, a droplet of ionic liquid (DEME-TSFI) was placed onto the device to cover the whole side gate and the rest of the sample. The device was then put into a refrigerator under high vacuum, and cooled down to ~220 K. We expedited this process to avoid the absorption of water or oxygen from the air and the consequent electrochemical reaction. All the electronic transport measurements were carried out in an Oxford Instruments Teslatron$^{TM}$ CF cryostat. A lock-in amplifier (Stanford Research 830) was used to measure the 4-probe resistance.

## ASSOCIATED CONTENT

## Supplementary information

The Supplementary information is available free of charge on the ACS Publications website at http://pubs.acs.org.

Characterization of $SnSe_2$ samples; Gate voltage $V_G$ dependence of $R_{sheet}$-$T$ and analysis of the effect of ionic liquid on sample with a high gate voltage; $T$-dependent magnetoresistance and $B$-dependent $V$-$I$ characteristics of $SnSe_2$; Analysis of $T$ dependence of $V$-$I$ and $R_{sheet}$-$T$ behaviors; temperature dependence of $B_{c2}$ in an out-of-plane and in-plane magnetic field; the modulation of magnetotransport under a perpendicular electric field.

AUTHOR INFORMATION

Corresponding Authors

E-mail: miao@nju.edu.cn (F. M.);

htyuan@nju.edu.cn (H. T. Y.);

sjliang@nju.edu.cn (S. J. L.).

Notes

The authors declare no competing financial interests.

Acknowledgements

We would like to thank Prof. Zhenhua Qiao, Yafei Ren, Dr. Peizhe Tang and Prof. Xiaoxiang Xi for fruitful discussions. This work was supported in part by the National Key Basic Research Program of China (2015CB921600, 2013CBA01603), the National Natural Science Foundation of China (61625402, 11374142, 61574076, 11474147), Fundamental Research Funds for the Central Universities and the Collaborative Innovation Center of Advanced Microstructures. This work was partially supported by the Department of Energy, Office of Basic Energy Sciences, Division of Materials Sciences and Engineering, under contract DE-AC02-76SF00515. F. M. thanks Dimojiu Sandubashao's group for experimental assistance and stimulating discussions.

# TOC Figure

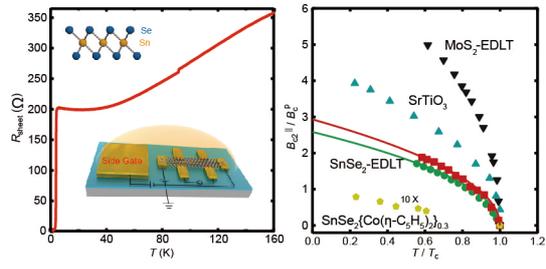